\documentclass[preprint2]{aastex}

\slugcomment{Draft version}
\shorttitle{$M_{\odot} of Massive Stars in Low-Z Galaxies}
\shortauthors{Tramper et al.}

\begin{document}

\title{On the Mass-Loss Rate of Massive Stars in the Low-Metallicity Galaxies IC~1613, WLM and NGC~3109\footnote{Based on VLT/X-Shooter observations under program 085D.0741.}}


\author{F. Tramper, H. Sana, A. de Koter and L. Kaper}
\affil{Astronomical Institute 'Anton Pannekoek', University of Amsterdam, Science Park 904, 1098 XH Amsterdam, The Netherlands}
\email{F.Tramper@uva.nl}


\begin{abstract}
We present a spectroscopic analysis of  VLT/X-Shooter observations of six O-type stars in the low-metallicity 
($Z \sim 1/7\,Z_{\odot}$) galaxies IC\,1613, WLM and NGC\,3109. The stellar and wind parameters of these
sources allow us, for the first time, to probe the mass-loss versus metallicity dependence of stellar winds
below that of the Small Magellanic Cloud (at $Z \sim 1/5\,Z_{\odot}$) by means of a modified wind momentum
versus luminosity diagram. The wind strengths that we obtain for the objects in WLM and NGC~3109 are unexpectedly high and do not agree with theoretical predictions. The objects in IC~1613 tend towards a higher than expected mass-loss rate, but remain consistent with predictions within their error bars. We discuss potential systematic uncertainties in the mass-loss determinations
to explain our results. However, if further scrutinization of these findings point towards an intrinsic cause
for this unexpected sub-SMC mass-loss behavior, implications would include a higher than anticipated number of
Wolf-Rayet stars and Ib/Ic supernovae in low-metallicity environments, but a reduced number of long-duration gamma-ray bursts produced through a single-star evolutionary channel.
\end{abstract}




\begin{deluxetable}{llrccccc}
\tablecolumns{8}
\tablewidth{0pc}
\tablecaption{Properties of the observed stars\label{tab:basic}}
\tablehead{\colhead{ID} & \colhead{R.A.} & \colhead{Dec.} & \colhead{$V$} & \colhead{$M_V$} & \colhead{Spectral} & \colhead{R.V.} & \colhead{12+log(O/H)} \\
\colhead{} & \colhead{(J2000)} & \colhead{(J2000)} & \colhead{} & \colhead{} & \colhead{Type} & \colhead{(km s$^{-1}$)} & \colhead{}}
\startdata
\textit{IC~1613} &&&&&&& \\
---A13 & 01 05 06.21 & +02 10 44.8 & 19.02 & -5.55 & O3 V((f))& $-$240 & 8.0\\
---A15 & 01 05 08.74 & +02 10 01.1 & 19.35 & -5.11 & O9.5 III & $-$240 & ---\\
---B11 & 01 04 43.82 & +02 06 46.1 & 18.68 & -5.84 & O9.5 I & $-$240 & ---\\
---C9 & 01 04 38.63 & +02 09 44.4 & 19.02 & -5.44 & O8 III((f)) & $-$265 & ---\\
\textit{WLM}&&&&&&& \\
---A11 & 00  01 59.97 & $-$15 28 19.2 & 18.40 & -6.35 & O9.7 Ia & $-$135 & --- \\
\textit{NGC~3109}&&&&&&&\\
---20 & 10 03 03.22 & $-$26 09 21.4 & 19.33 & -6.67 & O8 I & 407 & 7.8 \\
\enddata
\tablerefs{$V$-band magnitudes from 
\citet[][IC1613]{bres07};
\citet[][WLM]{bres06};
\citet[][NGC3109]{evan07}
}
\end{deluxetable}

\section{Introduction}\label{sec:intro}

The evolution of massive stars is greatly affected by the amount of mass and angular momentum lost during their lifetime.
Understanding the mechanisms responsible for these losses is a fundamental goal in stellar astrophysics. For
instance, mass-loss influences the characteristics of the supernova explosion with which a massive star ends its life, as well as the number of potential single-star progenitors of long-duration gamma-ray bursts \citep[e.g.][]{yoon05,woos06}. 

An important mass-loss mechanism, at least at galactic and Magellanic Cloud metallicities, is the transfer of momentum from photons to the atmospheric gas through line interactions, initiating and driving an outflow \citep[e.g., ][]{cast75,kudr00, vink01}. The strength of these radiation-driven winds therefore depends on the effective number of absorption lines, in particular near the photospheric flux maximum in the ultraviolet. It is dominated by the absorption of light through a copious amount of metallic ion lines that are present in this wavelength region. As a consequence, the mass-loss rate ($\dot{M}$) of stars hotter than 25000~K is predicted to scale with metallicity ($Z$) as $\dot{M} \propto Z^{0.69\pm 0.10}$ \citep{vink01}. \citet{moki07} showed that this prediction holds for early-type stars in the Galaxy ($Z=Z_{\odot}$), Large Magellanic Cloud (LMC; $Z=0.5\,Z_{\odot}$), and Small Magellanic Cloud (SMC; $Z=0.2\,Z_{\odot}$), yielding the empirical relation $\dot{M} \propto Z^{0.78 \pm 0.17}$. To date however, no observational constraints exist at sub-SMC metallicities.

For galactic stars more luminous than $~10^{5.8 }L_{\odot}$ (i.e. above the Humphreys-Davidson limit), 
episodic mass loss is expected to occur during the Luminous Blue Variable and/or Wolf-Rayet phase \citep[e.g., ][]{hump94, smit04}. Also, dust/pulsation-driven mass loss takes place in the red supergiant phase of lower luminosity objects \citep[e.g.,][]{hege97, yoon10}. However, if the {\em relative} contribution of line driving remains sizable at all $Z$, an important consequence of the theory of radiation-driven winds would be that low-metallicity massive stars loose less mass and angular momentum over their lifetime than their high-metallicity counterparts. 
At sub-SMC metallicities, single stars with a large initial rotational velocity may avoid envelope expansion and the associated efficient transfer of angular momentum from the core to the envelope, thus keeping a rapidly spinning core. This makes them potential progenitors of long-duration gamma-ray bursts \citep[e.g.][]{woos06}. 

In order to observationally constrain stellar and wind parameters at sub-SMC metallicities, we 
have used the X-Shooter spectrograph mounted on ESO's {\em Very Large Telescope} (VLT) UT2
\citep[][Vernet et al. 2011, in press]{dodo06} to secure intermediate-resolution spectra of some of the
most massive stars in three nearby dwarf galaxies, IC\,1613, WLM and NCG\,3109, each having a metallicity of $Z \approx 1/7 \ Z_{\odot}$. 
The throughput of the instrument, combined with the collecting area of an 8.2m telescope, allows us, for the first time, to perform a detailed quantitative spectral analysis at a resolution $R \sim 6000$-$9000$ 
and test the theory of line driving in O-type stars at a sub-SMC metallicity \citep[see also][]{herr11}.

With the aid of previous, low-resolution studies \citep{bres06,bres07,evan07}, we selected the six visually brightest O-type stars in these galaxies (four in IC~1613, one in both WLM and NGC~3109).
The low line-of-sight extinction towards these galaxies and their similar metallicity allow us, for this particular study,
to treat these stars as a group.

In Section~\ref{sec:observations} the observations and data reduction are described. Section~\ref{sec:modeling} discusses the method of analysis and Section~\ref{sec:results} presents the results. Finally, in Section~\ref{sec:discussion}, we discuss implications of our findings.

\begin{figure*}[!t]
%\epsscale{.80}
\centering
\includegraphics[scale=0.6,angle=90]{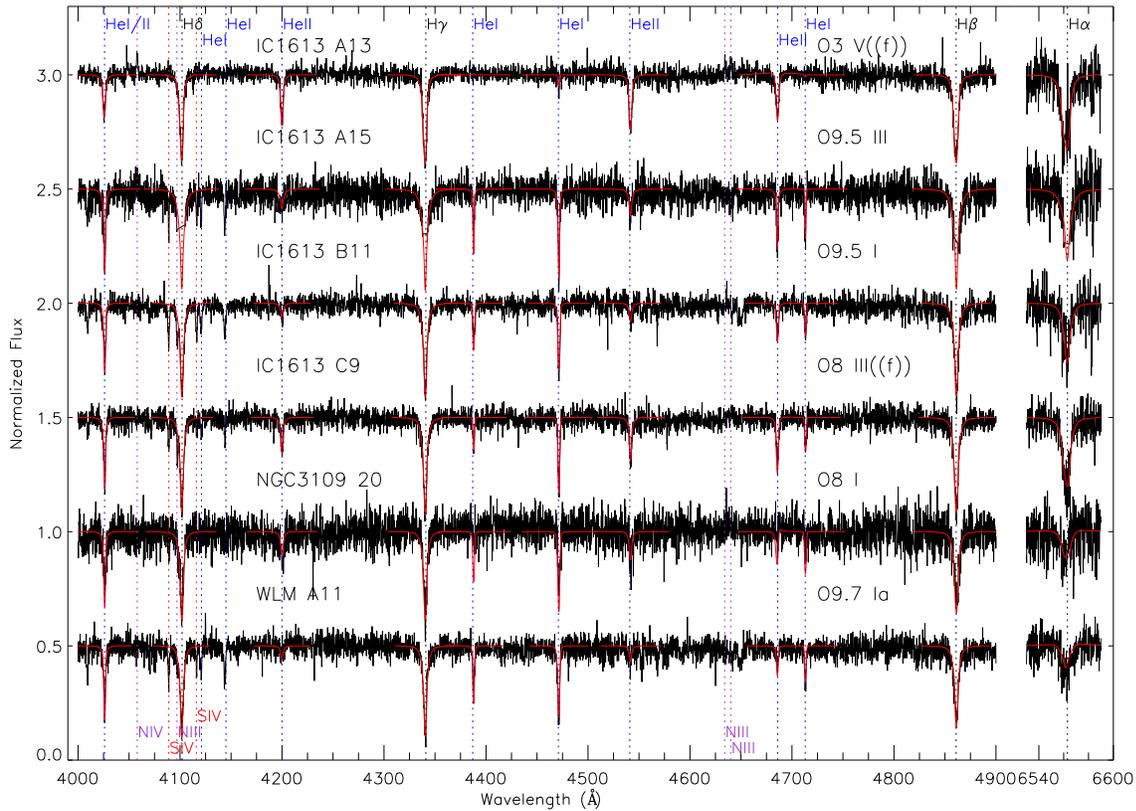}
%\plotone{atlas.eps}
\caption{Selected regions of the normalized X-Shooter spectra of the observed O-stars. Overplotted in red are the best-fit line profiles for the fitted lines.\label{fig:atlas}}
\end{figure*}

\section{Observations \& Data Reduction}\label{sec:observations}
The spectrum of NGC~3109-20 has been obtained during a X-Shooter GTO run on 2010 April 23, with an exposure time of 5x900s. In a successive GTO run from 2010 September 12 to 16, the stars in IC~1613 and WLM were observed, with an exposure time of 6x900s for all stars. All observations were obtained in nodding mode with a nod throw of 5", using a slit width of 0.8" for the UVB arm (300-550\,nm), 0.9" for the VIS arm (550-1020\,nm), and 0.9'' for the NIR arm (1000-2500\,nm), yielding a spectral resolving power of $R$=6200, 8800 and 5600, respectively. Conditions were clear or photometric with an average seeing below 0.8" in the R band. The moon illumination fraction was below 0.2.  Fundamental properties of the observed stars are listed in Table~\ref{tab:basic}.

\subsection{Data Reduction}
The data have been reduced using the X-Shooter pipeline v1.2.2 in physical model mode \citep{gold06,modi10}. The nodding mode usually allows for easy nebular subtraction. However, all sources except WLM-A11 show signs of variable strength of the nebular emission lines over the length of the nod throw, preventing the use of this method. For these stars the sky subtraction was done by using suitable regions of sky close to the object, resulting in a good nebular correction for all stars except IC~1613-A15 (see below). The signal-to-noise ratio (SNR) per resolution element ranges from 32 to 67 in the UVB arm and from 21 to 29 in the VIS arm. The detected flux in the NIR part of the spectrum was low (SNR $<$ 5), so we could not use this part of the spectrum in our analysis.

The nebular emission in IC~1613-A15 varies in radial velocity along the slit, causing residuals in the final spectrum. The regions affected by these residuals were removed from the final spectrum and not used in the atmosphere fit.

\begin{figure}[!t]
%\epsscale{.80}
\centering
\includegraphics[scale=0.98]{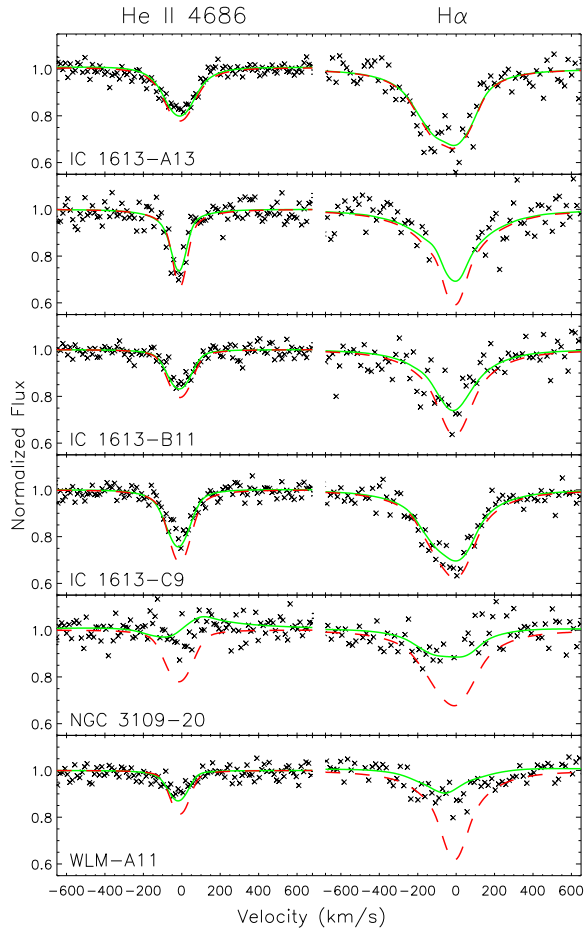}
%\plotone{atlas.eps}
\caption{Line profiles of the mass-loss sensitive diagnostic lines \ion{He}{2} 4686 and H$\alpha$. The best-fit line profiles are overplotted with solid lines. The dashed profiles indicate models with mass-loss rates predicted by radiation-driven wind theory. \label{fig:lines}}
\end{figure}

The resulting 1D spectra were normalized to the continuum. Figure~\ref{fig:atlas} presents the blue spectrum and H$\alpha$ region of all six stars. Figure~\ref{fig:lines} shows the profiles of the mass-loss sensitive \ion{He}{2} 4686 and H$\alpha$ lines in more detail. The spectral type of the stars was determined by comparing with a spectral atlas of O-type stars (Sana et al. 2011, in preparation), and for all but one agree with the type listed in the literature. The only change is that the spectral type of IC1613-A13 is refined from O3-4 V((f)) \citep{bres07} to O3 V((f)).

\subsection{Oxygen Abundance Determination}
In order to also provide an independent oxygen abundance measurement, a relative flux calibration was performed for the frames where nebular oxygen emission is present (IC~1613-A13 and NGC~3109-20), using spectra of photometric standard stars taken during each night of observations. To derive the oxygen abundance, we use the strong line method \citep{page79} with the updated calibration by \cite{pily05}, using the [\ion{O}{2}] $\lambda\lambda$ 3727,3729, [\ion{O}{3}] $\lambda\lambda$ 4959,5007 and H$\beta$ nebular emission lines. The resulting abundances are given in Table~\ref{tab:basic}, and although abundances derived with the strong line method can have uncertainties up to 0.8 dex \citep[see e.g.][]{bres09}, our measures agree well with the literature values of 12 + log(O/H) = $7.90\pm 0.08$ \citep[IC~1613;][]{bres07}, $7.80\pm 0.07$ \citep[WLM;][]{urba08} and $7.76\pm0.07$ \citep[NGC~3109;][]{evan07} derived from B- and A-type supergiants.

\section{Modeling}\label{sec:modeling}

The stellar properties and wind characteristics of the stars have been determined using an automated fitting method developed by \citet{moki05}. This method combines the non-LTE stellar atmosphere model \texttt{FASTWIND} \citep{puls05} with the generic fitting algorithm \texttt{PIKAIA} \citep{char95}. It allows for a fast and homogeneous analysis of our sample of stars using a selection of hydrogen, \ion{He}{1}, and \ion{He}{2} lines. 

The absolute visual magnitude ($M_V$) of the stars is needed as an input parameter. They have been determined using the $V$ magnitudes from Table~\ref{tab:basic} and a distance and reddening of $d=721$ kpc \citep{piet06} and $E(B-V)=0.025$ \citep{schl98} for IC~1613, $d=995$ kpc and $E(B-V)=0.08$ for WLM \citep{urba08}, and $d=1300$ kpc \citep{sosz06} and $E(B-V) = 0.14$ \citep{davi93} for NGC~3109. The resulting values of $M_V$ agree well with the values for their spectral type \citep{mart05}. The radial velocity (R.V.) has been measured by fitting a Gaussian profile to the hydrogen lines, and the obtained values are in agreement with the radial velocities of the host galaxies.

The fitting algorithm covers a large parameter space, fitting line profiles of the same 11 spectral lines as described by \cite{moki05} to determine the effective temperature ($T_{\mathrm{eff}}$), gravity ($g$), mass-loss rate ($\dot{M}$), surface helium abundance ($Y_{\mathrm{He}}$), depth-independent microturbulent velocity ($v_{\mathrm{tur}}$), and projected rotational velocity ($v_{\mathrm{rot}}\sin{i}$). The bolometric luminosity $L$ is derived by applying the bolometric correction to the absolute visual magnitude used as input. This luminosity, together with the obtained temperature, is then used to determine the radius ($R$) of the star. The terminal wind velocity ($v_{\infty}$) can not be constrained from the optical, but is related to the surface escape velocity $v_{\mathrm{esc}}$: $v_{\infty} = 2.6 \ v_{\mathrm{esc}}$ \citep[at $Z = Z_{\odot}$; ][]{lame95, kudr00}. The coefficient of the wind velocity structure $\beta$ has been fixed to 0.8 for the dwarfs, 0.9 for the giants, and 0.95 for the supergiants, conform theoretical predictions (Muijres et al. 2011, in preparation). 

\begin{deluxetable}{lcccllcccc}
\tablecolumns{10}
%\tablewidth{100pc}
\tablecaption{Best-fit parameters and derived properties of the observed stars\label{tab:fitting}}
\tablehead{\colhead{ID} & \colhead{$T_{\mathrm{eff}}$} & \colhead{log$g$} & \colhead{log$\dot{M}$} & \colhead{$Y_{\mathrm{He}}$} & \colhead{$v_{\mathrm{tur}}$} & \colhead{$v_{\mathrm{rot}}$sin $i$} & \colhead{$v_{\infty}$} & \colhead{log$L$} & \colhead{$R$}\\
\colhead{} & \colhead{\small{[kK]}} & \colhead{\small{[cm s$^{-2}$]}} & \colhead{\small{[$M_{\odot}$ yr$^{-1}$]}} & \colhead{} & \colhead{\small{[km s$^{-1}$]}} & \colhead{\small{[km s$^{-1}$]}} & \colhead{\small{[km s$^{-1}$]}} & \colhead{\small{[$L_{\odot}$]}} & \colhead{\small{[$R_{\odot}$]}}}
\startdata
\textit{IC~1613} &  &  &  &  &  &  &  &  & \\
---A13 & $47.6^{+4.73}_{-4.95}$ & $3.73^{+0.13}_{-0.22}$ & $-6.26^{+0.45}_{-0.50}$ & $0.32^{+0.04}_{-0.17}$ & $4^{+25}_{\downarrow}$ & $94^{+40}_{-28}$ & 1869 & 5.78 & 11.4\\
---A15 & $33.7^{+3.10}_{-2.65}$ & $3.76^{\uparrow}_{-0.39}$ & $-6.36^{+0.40}_{\downarrow}$ & $0.16^{+0.14}_{-0.10}$ & $9^{+17}_{\downarrow}$ & $34^{+42}_{-24}$ & 1971 & 5.24 & 12.2\\
---B11 & $31.3^{+2.00}_{-2.00}$ & $3.41^{+0.24}_{-0.21}$ & $-6.16^{+0.30}_{-1.30}$ & $0.13^{+0.11}_{-0.07}$ & $12^{+13}_{\downarrow}$ & $88^{+36}_{-28}$ & 1601 & 5.45 & 18.1\\
---C9 & $35.7^{+1.70}_{-1.85}$ & $3.58^{+0.19}_{-0.24}$ & $-6.26^{+0.25}_{-0.60}$ & $0.12^{+0.10}_{-0.04}$ & $15^{+8}_{\downarrow}$ & $72^{+28}_{-28}$ & 1697 & 5.43 & 13.6\\
\textit{WLM} & & & & & & & & & \\
---A11 & $29.7^{+2.45}_{-2.75}$ & $3.25^{+0.29}_{-0.19}$ & $-5.56^{+0.20}_{-0.30}$ & $0.11^{0.11}_{\downarrow}$ & $8^{+14}_{\downarrow}$ & $70^{+40}_{-36}$ & 1711 & 5.79 & 29.8\\
\textit{NGC~3109} & & & & & & & & & \\
---20 & $34.2^{+4.70}_{-3.05}$ & $3.48^{+0.37}_{-0.40}$ & $-5.41^{+0.25}_{-0.35}$ & $0.11^{\uparrow}_{\downarrow}$ & $16^{\uparrow}_{\downarrow}$ & $98^{+86}_{-72}$ & 2049 & 5.88 & 24.7\\
\enddata
\tablecomments{Arrows indicate upper or lower limits.}
\end{deluxetable}

Table~\ref{tab:fitting} presents the best-fit values and derived properties for the six stars. The values for $T_{\mathrm{eff}}$ are higher than those of their Galactic counterparts of the same spectral type \citep{mart05}, in agreement with their low metallicity \citep{moki04}. IC~1613-A13 shows an unusually high value for $Y_{\mathrm{He}}$. This is likely caused by the degeneracy between effective temperature and surface helium abundance at high temperatures (resulting from the loss of \ion{He}{1} as a diagnostic). The large uncertainties in the determination of the micro-turbulent velocity show that this parameter cannot be well constrained with our data.

The errors on the parameters have been analyzed by calculating the probability ($P = 1 - \Gamma \left( \chi^2/2, \nu/2 \right)$, where $\Gamma$ is the incomplete gamma function and $\nu$ the degrees of freedom) for all calculated models. Because $P$ is very sensitive to the value of the $\chi^2$, we normalize all $\chi^2$ values such that the best $\chi^2_{\mathrm{red}}$ is equal to 1, i.e. we assume that deviations of the original best $\chi^2_{\mathrm{red}}$ from unity are induced by under- or overestimated errorbars on the normalized flux. This approach is similar in spirit to using relative weighting in the $\chi^2$ merit function and to propagating the root-mean-square of the fit to scale the error bars \citep[e.g.,][]{pres86}. The uncertainties are obtained by considering the range of models which satisfy $P > 5\%$. These errors do not take into account uncertainties in the luminosity. However, as the mass-loss rate approximately scales with luminosity as $\dot{M} \propto L^{5/4}$, typical uncertainties in $L$ do not have a significant impact on our conclusions.

\section{Results \& Discussion}\label{sec:results}

The derived mass-loss rates are represented in the modified wind-momentum vs. luminosity diagram (WLD, Figure~\ref{fig:wld}). The modified wind momentum is defined as $D_{\mathrm{mom}} = \dot{M} v_{\infty} \sqrt{R/R_{\odot}}$. This quantity is ideally suited to study the $\dot{M}(Z)$ relation because it is almost independent of mass, and $v_{\infty}$ and $R$ are usually relatively well constrained. As the H$\alpha$ recombination line is essentially sensitive to the invariant wind-strength parameter $Q=\dot{M} / (R^{3/2}v_{\infty})$ that is inferred from the spectral analysis, $D$ (for given $T_{\mathrm{eff}}$) scales with $L$, making it less sensitive to uncertainties in the luminosity. As $v_{\infty}$ is expected to scale with metallicity \citep[$v_{\infty} \propto Z^{0.13}$;][]{leit92}, and the wind-strength parameter $Q$ is invariant, the derived mass-loss rate is subject to a similar scaling. This scaling of $\dot{M}$ and $v_\infty$ has been applied to the values given in Table~\ref{tab:fitting}.

Compared to theoretical expectations \citep[][$Z = 0.14 Z_{\odot}$ dashed line in Figure~\ref{fig:wld}]{vink01}, all stars except IC~1613-A13 tend toward a higher than predicted mass-loss rate for their metallicity. Compared to the empirical relations found by \cite{moki07}, our values are reminiscent of the values measured for the LMC. For IC~1613-A15 we could only obtain an upper limit due to the nebular contamination, and IC~1613-B11 and C9 could have a low enough mass-loss rate within the errors. WLM-A11 and NGC~3109-20 have a well defined mass-loss rate which is almost an order of magnitude too large for their metallicity.

\subsection{Systemetic Uncertainties in the Mass-Loss Determination}

In addition to random uncertainties, several sources of systematic uncertainties may affect our mass-loss determinations, for instance due to assumptions we have made. Here we discuss potential non-intrinsic causes for the high mass-loss rates that we derive.

The adopted values of the flow acceleration parameter $\beta$ could contribute to the high values of $D_{\mathrm{mom}}$. To fit the line profile, the wind density vs. velocity profile must be the same, causing a higher value of $\beta$ to give a lower mass-loss rate. If we underestimated $\beta$ the mass-loss rate will therefore be too high. However, even for extreme values of $\beta$, the correction will be at most a factor of two. The assumption that $v_{\infty} = 2.6 \ v_{\mathrm{esc}}$ introduces an uncertainty in $D_{\mathrm{mom}}$, but as errors in the terminal flow velocities are not expected to exceed 20-40\% \citep{groen89}, this effect is even smaller. 

\begin{figure*}[!t]
%\epsscale{.80}
\centering
\includegraphics[scale=0.6,angle=90]{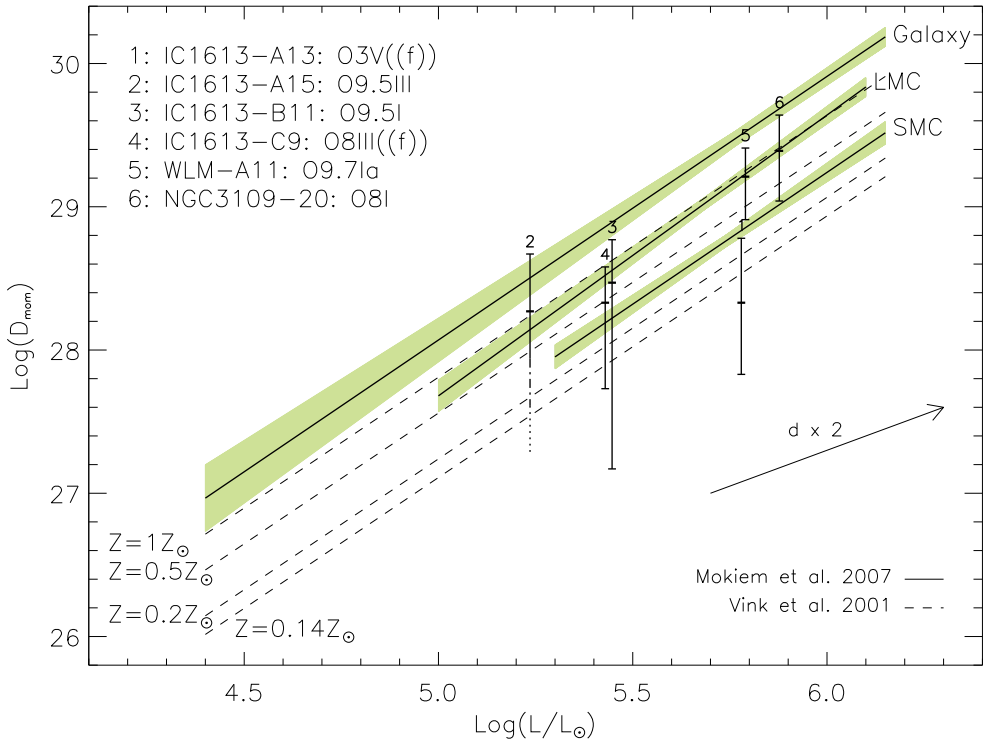}
%\plotone{atlas.eps}
\caption{Modified wind-momentum vs. luminosity diagram with our results. The dashed lines indicate the theoretical predictions of \cite{vink01}, and the empirical results from \cite{moki07} are represented by the shaded bars. The arrow illustrates the effect of an uncertainty in the distance on the location of points in the diagram.\label{fig:wld}}
\end{figure*}

Another source of uncertainty is the luminosity determination, both through extinction and distance uncertainties. However, as said before, the WLD is not very sensitive to uncertainties in $L$, as an increase in $L$ almost linearly increases the value of $D_{\mathrm{mom}}$. This effect is illustrated by the arrow in Figure~\ref{fig:wld}, indicating a change of a factor of two in distance. Furthermore, since the extinction towards all galaxies appears very low, and the absolute magnitude of the stars (Table~\ref{tab:basic}) agrees well with the values for their spectral type \citep{mart05}, we do not expect a significant error in the luminosity.

An underestimated metallicity could also be part of an explanation for the difference in empirical and predicted mass-loss rates. However, the various studies mentioned in Table~\ref{tab:basic} all agree on an oxygen abundance of 10-15\% solar. Our own determination of the oxygen abundance for IC~1613 and NGC~3109 also agrees with these values. Furthermore, earlier studies on the stellar populations of our target galaxies \citep{cole99,minn97,minn99} derived iron abundances of again 10-15\% solar.

Finally, multiplicity could play a role. Since the observed spectroscopic binary fraction in the Galaxy is approximately 50\% \citep{maso09,sana10}, it is possible that some of our sources are binaries. Binarity can influence the derived mass-loss rate by dilution or wind-wind collisions. In the case of a low-mass companion, the normalized spectrum will have shallower lines due to dilution effects. In close binaries with two massive stars, wind-wind collisions can cause emission in \ion{He}{2} $\lambda$4686 and H$\alpha$ if the wind-collision region is isothermal \citep[see, for example,][]{sana01}. However, we could not find indications
for massive companions in our data, which would cause higher luminosities, nor of radial velocities that differ from those of the parent galaxy. 

Because of the large distance to the host galaxies, it is possible that our targets are unresolved multiples instead of single stars (e.g., Hartoog et al. 2012, in preparation). With a seeing of 0.8", the smallest resolved area covers a distance ranging from 2.8 pc  for IC~1613 to 5 pc for NGC~3109, and can therefore easily harbor a typical open cluster. However, the observed luminosities are typical of stars of the corresponding spectral type \citep{mart05}, and we therefore do not see any strong indications for multiplicity.

\subsection{Possible Intrinsic Causes for the High Mass-Loss Rates}

There are several physical mechanisms that can increase the mass-loss rate or make it appear higher. A likely cause is the effect of wind clumping, which we do not take into account in our modeling. Clumping causes mass-loss indicators based on the average squared density $\left< \rho^2 \right>$, such as H$\alpha$, to have more emission than would be the case in an unclumped wind.
This causes the derived mass-loss rates to be overestimated by a factor of $1/\sqrt{f}$, with $f$ the mean value of the volume-filling factor in the line-forming region \citep{puls08}. 

However, by comparing our results to the relations found by \cite{moki07}, where clumping is also unaccounted for (causing the empirical mass-loss rate to be higher than the theoretical prediction), the rates we find for $Z \sim 0.14\,Z_{\odot}$ 
galaxies coincide with LMC values. So unless clumping behaves differently at sub-SMC metallicities, we would still expect the mass-loss rates to be located below the empirical SMC area in the WLD.

The slope and absolute scaling of the WLD is also anticipated to change at very low metallicities \citep{kudr02}. The expected effect, however, works in the opposite direction, and the mass-loss rates should be even lower. Furthermore, the slope-change is predicted to occur at a much lower metallicity ($Z \approx 10^{-3} Z_{\odot}$).

Fast rotation can also cause increased and asymmetric mass-loss \citep[e.g.,][]{maed00}. However, all of the projected rotational velocities we derive are low ($v\sin{i} < 100 \ \mathrm{km \ s}^{-1}$), and given that the probability that by chance all sources have their rotational axis pointed more or less in our direction is small, we do not expect this effect to play a major role in our sample. 

There could be other processes affecting the mass-loss process, possibly related to pulsations and magnetic fields, which both have been detected in O-type stars \citep[e.g.][]{henr99,dona09}. Pulsations are caused by the high sensitivity of opacity to temperature in the high-temperature low-density regions in O (super)giants, giving rise to $\kappa$-pulsations \citep[e.g.][]{igle92}. This is induced by the sudden appearance of a large number of same-shell transition iron lines, and is therefore dependent on the iron abundance. Thus it is expected that pulsations are less important at low metallicities \citep[][]{bara01}.

The increased opacity described above can give rise to small convective regions in the stellar envelope \citep{canti09}. \cite{canti11} showed that these convective regions can produce magnetic hot spots through the dynamo effect, which can be strong enough to influence the wind and possibly play a role in the wind clumping. This effect is again metallicity dependent, and is expected to play a lesser role and eventually disappear with decreasing metallicity \citep{canti09}.

\section{Conclusions and implications}\label{sec:discussion}

In this Letter, we have pointed out a discrepancy between the observed and predicted mass-loss rates from massive stars in low-metallicity environments, and discussed possible explanations. %However, our current sample of stars is too limited to draw firm conclusions, and more observations are needed to confirm our results. Planned X-Shooter observations of four more stars in IC~1613 are a first step towards this goal.

A potential violation of the expected metallicity scaling of the radiation-driven mass-loss rate at 
$Z \sim 1/7 \ \,Z_{\odot}$, resulting in higher than expected mass-loss rates, would have far-reaching implications.
Firstly, one expects low-$Z$ O-type stars to suffer more from spin-down through angular momentum loss in the stellar wind. 
Consequently, the rotational mixing efficiency may be reduced, leading, for instance, to a more modest nitrogen enrichment
than currently thought \citep[][]{brott11}. 

Secondly, one anticipates the single O star population in low metallicity environments to produce more observationally 
identifiable Wolf-Rayet stars, being the successors of O-type stars having their outer envelope stripped by mass-loss, and therefore
increased number of Ib and, potentially, Ic supernovae. The single-star channel would, however, produce less progenitors of long-duration gamma-ray bursts, as the stars
loose more angular momentum by their outflow. 

Finally, if the larger than expected wind strength at $Z \sim 1/7 Z_{\odot}$ persists to extremely low metallicities (though at present the driving mechanism is unknown), stellar winds
will impact the evolution of massive Population III stars and the chemical enrichment of the
intergalactic medium of the early Universe.

\acknowledgments We thank the referee, dr. Kudritzki, for his useful comments, dr. Martayan for his support during the observations, and O. Hartoog for the interesting discussions.


%% in parentheses, after the keyword, but they will not be verified.
{\it Facilities:} \facility{VLT:Kueyen (X-Shooter)}

\clearpage

\end{document}